\documentstyle[pra,aps,epsfig]{revtex}

\begin{document}
\title{\bf Coherence of trapped one-dimensional (quasi-)condensates  \\ and continuous 
atom lasers in waveguides}

\author{N.P.\ Proukakis}
\address{
Department of Physics, University of Durham, South Road, Durham DH1 3LE, United 
Kingdom \\
\mbox{}$^{\star}$E-mail: n.p.proukakis@durham.ac.uk}

\maketitle

\begin{abstract}

We present a model for a continuous atom laser beam in a one-dimensional
waveguide. The beam is 
formed by 
continuous raman outcoupling of a trapped one-dimensional (quasi-)condensate,
 which is created by imposing a tightly confining transverse optical
potential on a three-dimensional magnetically trapped ultracold
thermal gas. The trapped (quasi-)condensate is modelled by a stochastic Langevin
equation, in which pumping arises naturally from  the surrounding three-dimensional 
thermal cloud, which
acts as its heat bath. 
When the outcoupling is turned on, such pumping leads to continuous replenishing 
of the (quasi-)condensate, and thus to a steady-state operation 
for the atom laser, which is described by a nonlinear Schr\"{o}dinger equation, 
coupled to 
the Langevin equation for the (quasi-)condensate. This model is used to
study the temporal and spatial evolution of the 
coherence length of the (quasi-)condensate and the atom laser over the 
entire system.\\

\vspace{2.0cm}


\end{abstract}

%

\section{Introduction}


An atom laser is a device which emits an intense, directed
 coherent atomic beam in much the
same way as a conventional (optical) laser emits a beam of coherent photons.
Interest in such `devices', whose potential applications include 
fields such as precision metrology, atom holography and 
nanotechnology, arose after the experimental realization of
Bose-Einstein condensation (BEC) in trapped atomic alkali gases 
\cite{BEC_Exp1,BEC_Exp2,BEC_Exp3}.
Although BEC need not necessarily be a prerequisite for the production of an atom laser,
it was soon realized that a coherent atomic beam could be generated by the coherent 
extraction of a
trapped BEC, thus leading to the first
demonstration of a `rudimentary atom laser beam' in 1997 
\cite{AtomLaser_Exp1a,AtomLaser_Exp1b}.
In this pioneering experiment, atoms from a BEC trapped in a particular magnetic
state were coherently extracted
 by means of pulsed radio-frequency (rf) transitions to a magnetically untrapped state;
the outcoupled atoms then fell  under gravity. Since then, a number of experimental 
groups have demonstrated the ability to produce pulsed 
\cite{AtomLaser_Pulsed2,AtomLaser_Pulsed3},
 or quasi-continuous 
\cite{AtomLaser_CW1a,AtomLaser_CW1b,AtomLaser_CW2,AtomLaser_CW3,AtomLaser_CW4} 
atom
lasers, with atoms usually extracted from the magnetically trapped state via rf
or Raman transitions (although gravity-induced tunnelling from an optically 
trapped BEC has also been demonstrated as an outcoupling mechanism).
The outcoupled atoms propagate away from the centre of the trap with an
accelerated motion due to gravity, or at a constant speed due to the momentum
imparted to the atoms by the two-photon Raman transitions. 
Raman outcoupling has the additional advantage of offering good spatial
selectivity within the condensate.
The removal of trapped atoms
depletes the condensate and currently sets a very tight constraint on the duration
of atom laser operation.
Significant steps to overcome this limitation were recently carried out at MIT,
where it was demonstrated that the time required for the creation of a reasonably-sized BEC
can be significantly smaller than its lifetime \cite{Zora}. This has enabled
researchers to make a second BEC, merge it with an existing one, thus 
demonstrating the first 
real-time replenishing of a condensate. However, at the time of writing this paper,
it is still unkown how such merging affects the coherence properties of the original
condensate. All above discussion focuses on the production of an atom laser beam based on
the coherent extraction a of ground-state
BEC, and an interesting topic meriting further discussion is the possibility of
creation of coherent atomic beams with different spatial modes, based on the
coherent extraction of a non-ground-state BEC \cite{Yukalov}. 
Finally, one should mention an alternative approach 
currently underway experimentally \cite{Guery}, which aims to
create a 
continuous atom laser by evaporative cooling in a magnetic waveguide \cite{Yvan}.

The coherence properties of trapped condensates 
have been studied experimentally using a variety of techniques. 
Spatial (equilibrium) properties have been studied by interfering two 
independent BEC's \cite{AtomLaser_Exp1b},
by studying two- \cite{SpatialCoh_TwoBody} 
and three-body interactions \cite{SpatialCoh_ThreeBody},
by spectroscopic techniques \cite{Spatial_Coh1},
and by interfering atoms extracted from different points within a single condensate
\cite{AtomLaser_CW1b}, or at different times \cite{Spatial_Coh2,Temp_Coh2}.
The temporal coherence of an atom laser beam extracted from a pre-formed BEC
 has been measured in \cite{Temp_Coh1}, thus also yielding an upper limit for temporal
phase fluctuations within the condensate.
On the theoretical side, the coherence properties of atom lasers have been studied
in numerous publications, and an excellent review of the early literature on this
topic (e.g. based on rate equations \cite{Rate_Eqns}) 
can be found in  \cite{Atom_Laser_Theory}.
Our theoretical understanding of these issues has further improved following 
the successful analysis of a recent experiment by means of the Gross-Pitaevskii equation
\cite{Temp_Coh2} and an in-depth discussion
of output coupling based on Raman transitions \cite{Raman_Theory}.
However, such treatments are aimed at describing the coherence properties of an atom 
laser as it is extracted from a pre-formed (pure) condensate.
These treatments hence suffer from two limitations: Firstly, they completely
ignore the effect of thermal atoms which are inevitably present in the system 
and can lead to additional phase diffusion of the atom laser beam \cite{Japha,Janne}.
Perhaps more significantly for our current discussion, such treatments do not
consider a re-pumping mechanism into the condensate, to compensate for the
outcoupled atoms.
These treatments are therefore not well suited for describing steady-state atom laser
operation. The latter necessitates a mechanism pumping atoms into the condensate, and
this is usually modelled phenomenologically \cite{Phen1,Phen2,Phen3}.
In particular, in recent discussions  \cite{Savage_AL_PRA,Savage_AL_PRL}, both 
trapped and outcoupled components are described
by coupled nonlinear Schr\"{o}dinger equations, with the trapped component
additionally pumped from a thermal cloud, in such a manner, that the condensate is
continually replenished during the outcoupling.

However, such pumping is assumed to be coherent, such that
 any atom falling into the magnetic trap
becomes immediately part of the condensate, whereas we believe that a realistic
pumping mechanism would, in general, introduce an additional phase diffusion of the 
condensate, as a result of the newly-pumped atoms.
To account for this additional feature, one must combine the description of 
condensate formation based on quantum kinetic 
theory, with the process of coherent outcoupling. 
It is the aim of this paper to develop such
a description and perform a qualitative and quantitative analysis of the coherence
properties of a continuous one-dimensional atom laser. 
The results presented here are somewhat preliminary and will be further investigated
in a subsequent publication.
The important new feature of our model is that
 pumping arises naturally from the
non-equilibrium treatment of the coupled dynamics of condensed and
thermal atoms in a trap \cite{Stoof_Booklet}. 
This gives rise to a Langevin equation for the growth of a condensate
in contact with a heat bath \cite{Stoof_Noisy}
(note that such an approach differs considerably from previous stochastic treatments applied
to similar issues \cite{Stoch1,Stoch2}).
The outcoupled atoms  are then described by a nonlinear 
Schr\"{o}dinger equation, which is coupled to the stochastic condensate field
by (externally-induced) Raman coupling. 
The fluctuations inherent in the trapped component are then
passed onto the atom laser beam, whose
coherence properties we can thus determine.

This work focuses entirely on one dimensional systems, which 
are known to suffer from enhanced 
phase fluctuations and possess no off-diagonal long-range order. In particular,
if such phase variations occur on length scales
considerably smaller that the size of the system, then the system is known 
as a quasi-condensate \cite{Popov,Quasi_BEC}. 
This significant difference in the coherence properties of one-dimensional systems
from the three-dimensional case usually encountered in current atom laser
experiments indicates that our approach will yield
estimates for the atom laser coherence which will be strongly
dependent on the coherence properties of the original trapped (quasi-)condensate,
and will thus, in general, not be directly applicable to current experiments.
For this reason, our current work does not focus on the `absolute' coherence 
of the atom laser, but rather on the dependence of the coherence
properties of the output beam on the outcoupling procedure
(given a particular coherence of the trapped system), and the change
in the coherence properties of the trapped component due to the outcoupling.
To some extent, such conclusions should also hold qualitatively for the three-dimensional
systems. In this paper we discuss a continuous raman
output coupling scheme to a magnetically untrapped state, with the atoms leaving
the `interaction region' at the centre of the trap by means of the momentum
imparted to them by the Raman lasers \cite{AtomLaser_CW2}.
Throughout this work we assume that the entire system (condensate plus atom laser)
remains kinematically one-dimensional, which is equivalent to an atom laser
propagating in a transversely very tightly confining waveguide \cite{Hannover_Wvguide}. 
For technical reasons, the preliminary results presented in this paper are limited
to the region of quasi-condensation, with parameters closely resembling those
of a recent one-dimensional experiment \cite{MIT_1D}. 
The  case of (almost) true condensation, and a more in-depth theoretical interpretation 
of the predictions of our model, will be discussed in a separate publication.

This paper is structured as follows. Sec. 2 discusses in detail the theoretical framework
used to study the coherence properties of a (quasi-)condensate in a trap and the atom 
laser beam. In Sec. 3 we present results for the 
coherence length of the trapped component, prior to the outcoupling (Sec. 3.1) and
after the outcoupling has been  turned on (Sec. 3.2). Sec. 4 discusses the 
 coherence length for the outcoupled component, and compares its behaviour
at steady state to that of
the trapped (quasi-)condensate. In particular, we discuss explicitly how the
coherence length of the atom laser beam is modified by changing either the
magnitude, or the phase of the external electromagnetic field 
(corresponding to the momentum imparted to the outcoupled atoms) 
producing the outcoupling.
 Finally, Sec. 5 presents some concluding remarks.

\section{Theoretical Model}


In this section we present the theoretical framework to be used in the rest of the paper.
Firstly, we describe the one-dimensional 
(quasi-)condensate in the trap.

\subsection{Trapped Component Only}


Consider an ultracold 
magnetically trapped three-dimensional gas above the critical point, on which
a laser beam is applied in such a manner that it
provides an additional tightly-confining optical potential along two of the directions, in a manner similar to a recent experiment \cite{MIT_Dimple}.
The laser beam can be arranged in such a manner that the
potential `dimple' it creates transversely to the waveguide has
 only one bound state
below the chemical potential of the three-dimensional thermal cloud. In this case,
 the motion
of the system becomes `frozen out' along the transverse directions. The system
thus becomes kinematically one-dimensional, but remains in  contact with the 
three-dimensional thermal cloud, which acts as its 
heat bath. The dynamics of the order 
parameter  $\Phi(z,t)$ in the dimple is thus governed 
by~\cite{Stoof_Booklet,Stoof_Noisy}
\begin{equation} 
i \hbar \frac{ \partial \Phi(z,t) }{ \partial t} 
=  \Bigg[ - \frac {\hbar^{2} \nabla^{2} }{2m} + V^{\rm ext}(z) - \mu 
- iR(z,t)  
+ g |\Phi(z,t)|^{2} \Bigg] \Phi(z,t) + \eta(z,t)\;,
\label{lang}
\end{equation}
where the external trapping potential in the weakly-confining
direction is given by $V^{\rm ext}(z)=m \omega_{z}^{2} z^{2} /2$,
and $\mu$ is the effective chemical potential of the one-dimensional system. 
The one-dimensional coupling constant $g$ is given by 
$g=4\pi \hbar^2 \kappa /m$, where $\kappa$ corresponds to a one-dimensional
scattering length; this is related to the usual 
three-dimensional scattering length 
$a_{3D}$ by $\kappa = a_{3D} / 2 \pi l_{\perp}^{2}$, where 
$l_{\perp}=\sqrt{(\hbar / m \omega_{\perp})}$ is the harmonic oscillator length
of the axially symmetric trap in the direction perpendicular to the waveguide axis z.
Note that $\Phi(z,t)$ corresponds to the {\bf entire} field in the dimple, and not
simply to the condensed component. The role of the thermal component is evident in the
contributions $iR(z,t)$ and  $\eta(z,t)$.
Physically, the function $iR(z,t)$ describes the
pumping of the one-dimensional gas from the surrounding reservoir, 
and $\eta(z,t)$ corresponds to the associated noise with Gaussian 
correlations.
In the classical field approximation~\cite{Stoof_Booklet,Stoof_Noisy}
 (which is well-satisfied in current experiments), 
these are given by 
\begin{equation}
iR(z,t)=-{\beta\over4}\hbar\Sigma^{\rm K}(z)
\left(
-{\hbar^2\nabla^2\over2m}+V^{\rm ext}(z)-\mu+g|\Phi(z,t)|^2
\right)\;, 
\end{equation}
\begin{equation}
\langle
\eta^*(z,t)\eta(z^\prime,t^{\prime})
\rangle
={i\hbar^2\over2}\Sigma^{\rm K}(z)\delta(z-z^\prime)\delta(t-t^{\prime})\;,
\end{equation}
where $\langle...\rangle$ denotes averaging over the realizations 
of the noise $\eta(z,t)$, and $\beta=1/(k_{B}T)$ as usual.
Both above quantities depend on the Keldysh
self-energy $\hbar\Sigma^{K}(z)$ which arises physically from collisions that scatter
an atom out of, or into the `heat bath', and is given by
\begin{eqnarray}
\hbar\Sigma^{K}(z) & = & - 4 i \frac{g^{2}}{(2 \pi)^{5} \hbar^{6}}
\int d p_{1}  \int d p_{2} \int d p_{3}
\delta(p_{1}-p_{2}-p_{3})
\delta(\epsilon_{1}-\epsilon_{2}-\epsilon_{3}+V_{ext}) \nonumber \\
& & \times (N(\epsilon_{1})+1)N(\epsilon_{2})N(\epsilon_{3})
\end{eqnarray}
where $\epsilon_{i}=(p_{i}^{2}/2m)+V_{ext}(z_{i})$ and 
$N(\epsilon_{i})=[exp(\beta (\epsilon_{i}-\mu)) -1]^{-1}$
is the Bose occupation factor.
The numerical techniques employed are discussed in 
Ref.~\cite{Stoof_Noisy}, where it was also shown that 
the above expressions guarantee that
the trapped gas relaxes to the correct equilibrium, as ensured by the 
fluctuation-dissipation theorem. To simplify the numerics, 
the noncondensed part in the dimple is here
allowed to relax to the ``classical'' value 
$N(\epsilon)=[\beta ( \epsilon- \mu )]^{-1}$.

\newpage

Fig. 1 shows the evolution of the total density profile in the one-dimensional
trap at different relaxation times (after turning on the optical dimple), 
for a sodium (quasi-)condensate at $T=200 nK$
and with trapping frequencies taken from a recent experiment \cite{MIT_1D}.
The effect of the thermal contributions is obvious, since the equilibrium
density profile differs considerably from the corresponding (zero-temperature) Thomas-Fermi 
solution 
 obtained in the absence of noise.

\begin{figure}
\centerline{\psfig{figure=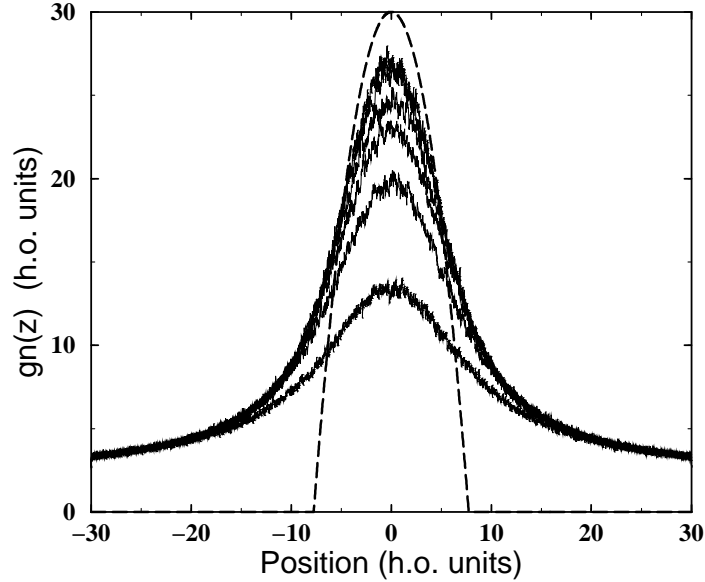,height=8.0cm}}
\caption{Total density profile in the dimple after variable relaxation times,
 with $t=0$ corresponding to the time when the optical trap is
switched on.
From bottom to top (noisy curves):
$t_{0}/\tau = $ 4.8, 9.6, 14.4, 19.2, 28.8 and 38.4 (corresponding to final
relaxed value which is indistinguishable from
that of 28.8).  In this
and subsequent figures, time and length
 are scaled to harmonic oscillator units, respectively  
$\tau = \omega_{z}^{-1} = 45.5 ms$ and $l_{z} = \sqrt{(\hbar / m \omega)} = 11.2 \mu m$. 
The dashed line shows the corresponding Thomas-Fermi profile in the absence of thermal component (i.e. without noise),
with the zero-temperature Thomas-Fermi radius acquiring a value $R_{TF}=7.74 l_{z}$.
Throughout this paper, we use the magnetic trapping frequency 
$\omega_{z} = 2 \pi \times 3.5$ Hz, the transverse (optical) confining potential 
$\omega_{\perp} = 2 \pi \times 360$ Hz, and an effective one-dimensional
 chemical potential $\mu = 30 \hbar \omega_{z}$.
For $^{23}Na$, $m=3.817 \times 10^{-26} kg$ and the three-dimensional 
scattering length is $a_{3D} = 2.75 nm$.
Throughout this paper, the surrounding thermal cloud is at a 
temperature $T = 200 nK$. The discretization
used is $dx= 0.05 l_{z} $ and $dt= 0.0008 \tau$, which we have checked 
is sufficient to produce correctly converged results.}
\end{figure}

\subsection{Coupled Evolution of Trapped and Outcoupled Components}


Having discussed how a one-dimensional (quasi-)condensate is produced, 
we next describe our model for
a one-dimensional atom laser. After allowing the trapped component to relax to its
equilibrium value, the outcoupling mechanism is turned on, leading to a continuous
depletion of the (quasi-)condensate. Nonetheless, the presence of the additional
 pumping mechanism from the surrounding thermal cloud (which acts as its heat bath), 
leads to a steady-state population of the trapped component. 
An implicit crucial assumption made here is that the
heat bath remains unaffected as it pumps atoms into the (quasi-)condensate, to 
compensate for the atoms lost from the trap due to outcoupling.
This would indeed be the case (at least for short times) in the limit of
very slow outcoupling and large initial thermal cloud. In our model, we envisage that
 the ultracold thermal cloud is continually replenished
at a rate comparable to that of outcoupling. Then,
our model is expected to be valid as long as the residual oscillations
 in the heat
bath atom number (which indirectly affect the coherence of the trapped component
and of the  atom laser) are kept arbitrarily small, which we believe lies within
current experimental capabilities.
As noted earlier, we further assume that our entire
system remains kinematically one-dimensional, which requires the optical dimple to 
extend over the entire range of our atom laser.

To model the raman outcoupling scheme, we follow the usual procedure
of coupling the order parameter of the trapped system to a nonlinear Schr\"{o}dinger
equation for the outcoupled component. Such an approach has been already discussed
in the literature \cite{Atom_Laser_Theory,Savage_AL_PRA,Savage_AL_PRL}, 
by making the further crucial assumption that the trapped
component can be adequately described by a nonlinear 
Schr\"{o}dinger equation with phenomenological pumping.
Our treatment improves on such earlier attempts in that we explicitly include the 
effect of the thermal cloud on the traped component 
by means of a non-equilibrium theory based on the
Schwinger-Keldysh formalism \cite{Stoof_Booklet}. Hence, in our treatment pumping 
arises as a result of the relaxation of the
system into its new confining potential.

The coupled evolution of the trapped $\Phi_{1}(z,t)$ and outcoupled $\Phi_{2}(z,t)$
components is thus
determined by
\begin{eqnarray}\nonumber
i \hbar \frac{ \partial \Phi_{1}(z,t) }{ \partial t} 
&= & \Bigg[ - \frac {\hbar^{2} \nabla^{2} }{2m} + V_{1}^{\rm ext}(z) - \mu + \delta
- iR(z,t)  
+ g_{11} |\Phi_{1}(z,t)|^{2} + g_{12} |\Phi_{2}(z,t)|^{2} \Bigg] \Phi_{1}(z,t) \\ &&
+ \Omega(z,t) \Phi_{2}(z,t) + \eta(z,t)
\label{lang}
\end{eqnarray}
\begin{eqnarray}\nonumber
i \hbar \frac{ \partial \Phi_{2}(z,t) }{ \partial t} 
& = & \Bigg[ - \frac {\hbar^{2} \nabla^{2} }{2m} + V_{2}^{\rm ext}(z) - \mu
+ g_{22} |\Phi_{2}(z,t)|^{2} + g_{12} |\Phi_{1}(z,t)|^{2} \Bigg] \Phi_{2}(z,t) \\ &&
+ \Omega^{*}(z,t) \Phi_{1}(z,t)\;.
\label{lang2}
\end{eqnarray}
Here $g_{ii}$ correspond  to the one-dimensional self-interactions of each
component, whereas $g_{12}$ is the mean-field interaction between the two components.
For simplicity, we have taken $g_{11}=g_{22}=g_{12}=g$ (defined earlier).
The two components  are coupled by the application of an external electromagnetic
field via
$\Omega(z,t) = \Omega_{0} e^{ikz}$, where $\Omega_{0}$ the strength of the Raman coupling
and $\hbar k$ the momentum imparted to the outcoupled atoms. 
Our description is limited to a one-dimensional atom laser 
in a {\bf horizontal} waveguide, and thus ignores gravity.
For $^{23}Na$, the trapped atoms are assumed to be in the 
$|F,m_F \rangle = |1,-1 \rangle $ state 
(for which $V_{1}^{\rm ext}(z)=m \omega_{z}^{2} z^{2} /2$), whereas for
simplicity we have chosen to outcouple the atoms to the untrapped 
$|F,m_F \rangle = |1,0 \rangle $ state 
(for which  $V_{2}^{\rm ext}(z) = 0$). 
All our simulations are performed at zero detuning
($\delta=0$) such that the outcoupling, 
which is peaked precisely at $z=0$, occurs only in a very narrow region around 
the centre of
the trap. An illustrative discussion of the effect of changing the detuning can be
found in \cite{Choi_Outcoupling}.

Fig. 2 shows the density profiles of the trapped and outcoupled components at
steady state, for two different coupling parameters $\Omega_{0}$. We see clearly that
the momentum kick imparted to the outcoupled atoms, leads to their propagation in the
negative z-direction (note that absorbing boundaries have been used at the edges
of our computational grid). 
Due to the mean field interaction  between trapped and outcoupled
components, such directional motion of the outcoupled atoms creates an asymmetry in
the profile of the trapped atoms, which is more pronounced in the case of large $\Omega_{0}$.
Since the outcoupling is primarily from the trap centre ($\delta = 0$), this leads to
a ``double-peak'' structure for the trapped atom density profile.

\begin{figure}
\centerline{\psfig{figure=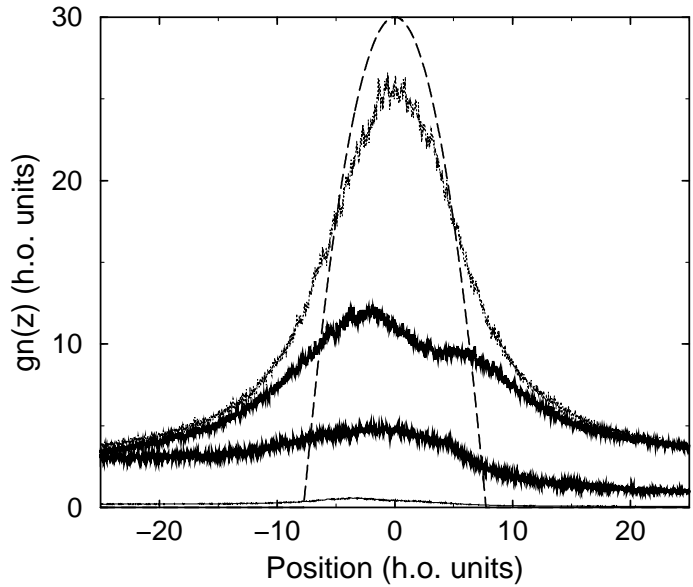,height=8.0cm}}
\caption{Density profiles of trapped atoms (top two noisy curves) and outcoupled atoms 
(bottom two noisy curves) at steady state ($t_{0}/\tau=38.4$),
in the limit of (i) `weak outcoupling': $\Omega_{0}=0.5 \hbar \omega_{z}$ (thin noisy curves)
and
(ii) `strong outcoupling': $\Omega_{0}=5 \hbar \omega_{z}$ (thick noisy curves),
 for $\delta=0$. Also shown is the Thomas-Fermi profile (dashed line).
Note that  the equilibrium quasi-condensate profile prior to
outcoupling essentially overlaps with the profile of the trapped atoms in the
limit of `weak outcoupling' shown here.
In both cases the momentum imparted to the atoms is $k = -5 l_{z}^{-1}$, 
corresponding to an
atomic speed $v = (\hbar k /m) \sim 1.2 mm/s$ along the negative z-axis.}
\end{figure}

Having discussed our model and its predictions for the density profiles we now turn
 to the discussion of the coherence properties, 
which we investigate in terms of an appropriately
defined coherence length.

\subsection{Definition of Coherence Length}


The coherence properties of the system can be inferred from the behaviour of the
normalized first-order off-diagonal correlation function defined by
\begin{equation}
g^{(1)}(z_{0},t_{0};z,t) = \frac{ \langle \Phi_{\alpha}^{*}(z_{0},t_{0}) \Phi_{\alpha}(z,t) 
\rangle }
{ \sqrt{ \langle |\Phi_{\alpha}(z_{0},t_{0})|^{2} \rangle \langle |\Phi_{\alpha}(z,t)|^{2} 
\rangle }}\;,
\end{equation}
where $\langle...\rangle$ denotes averaging over noise realizations, 
and $\alpha=1,2$ denote respectively the trapped and outcoupled components.
Such a quantity can be calculated by numerical autocorrelation measurements.
Fig. 3 shows the position dependence of the equal-time correlation function 
$g^{(1)}(z_{0},t_{0};z,t_{0})$ for the trapped component at the trap 
centre ($z_{0}=0$), once it has relaxed to
its equilibrium value in the absence of outcoupling. 
The observed rapid decay of the correlation function over the size of the system is 
clear evidence of quasi-condensation, as discussed in \cite{Low_D_Paper}. 
Since this
 closely resembles an exponential \cite{Exp_Decay,Castin_AL}, we have 
chosen to define the coherence length, $l_{coh}$ as the value of $z$ for which this 
function decays to $(1/e)$
(more generally $g^{(1)}(z_{0},t_{0};(z_{0}+l_{coh}),t_{0})=(1/e)$. 
In this paper we will investigate the behaviour of $l_{coh}$ 
(i) as a function of $z_{0}$
for both trapped and outcoupled components, at fixed relaxation time $t_{0}$, and 
(ii) as a function of relaxation time $t_{0}$, for fixed position $z_{0}$.
Note that by relaxation time we define the time elapsed from the moment that the
tightly confining transverse potential is turned on.

\begin{figure}
\centerline{\psfig{figure=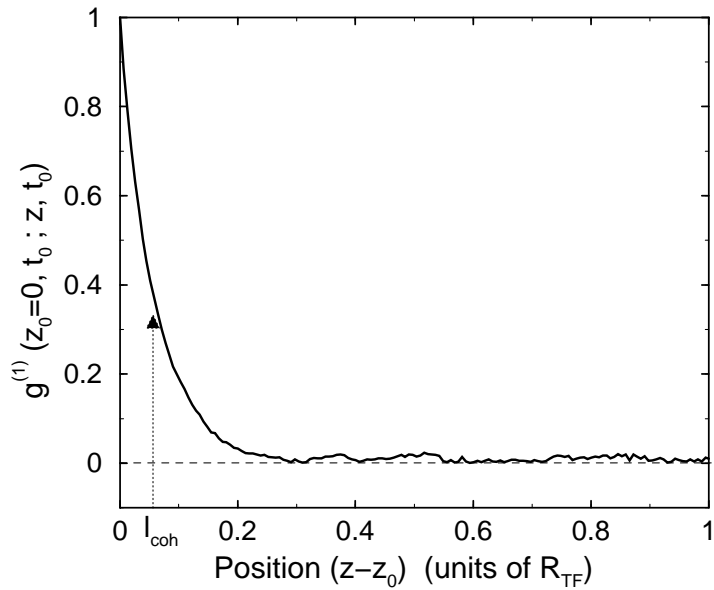,height=8.0cm}}
\caption{Density profiles of trapped atoms (top two noisy curves) and outcoupled atoms 
(bottom two noisy curves) at steady state ($t_{0}/\tau=38.4$),
in the limit of (i) `weak outcoupling': $\Omega_{0}=0.5 \hbar \omega_{z}$ (thin noisy curves)
and
(ii) `strong outcoupling': $\Omega_{0}=5 \hbar \omega_{z}$ (thick noisy curves),
 for $\delta=0$. Also shown is the Thomas-Fermi profile (dashed line).
Note that  the equilibrium quasi-condensate profile prior to
outcoupling essentially overlaps with the profile of the trapped atoms in the
limit of `weak outcoupling' shown here.
In both cases the momentum imparted to the atoms is $k = -5 l_{z}^{-1}$, 
corresponding to an
atomic speed $v = (\hbar k /m) \sim 1.2 mm/s$ along the negative z-axis.}
\end{figure}

\newpage
\section{Coherence Length of Trapped System}


\subsection{Without Outcoupling}


Initially we discuss the  relaxation of the trapped component 
to its equilibrium value in the dimple prior to the outcoupling.
 Fig. 4(a) shows the
temporal evolution of the coherence length at various positions in the trap. It is
seen clearly that $l_{coh}$ grows significantly at the trap centre and remains mostly
unaffected at the edges of the trap (where there are only thermal atoms present).
The dependence of $l_{coh}$ on position in the trap at various times can also be
seen in Fig. 4(b), which confirms the anticipated result that
the coherence length is largest at the trap centre (where the quasi-condensate density
is maximum). 

\begin{figure}
\centerline{ \psfig{figure=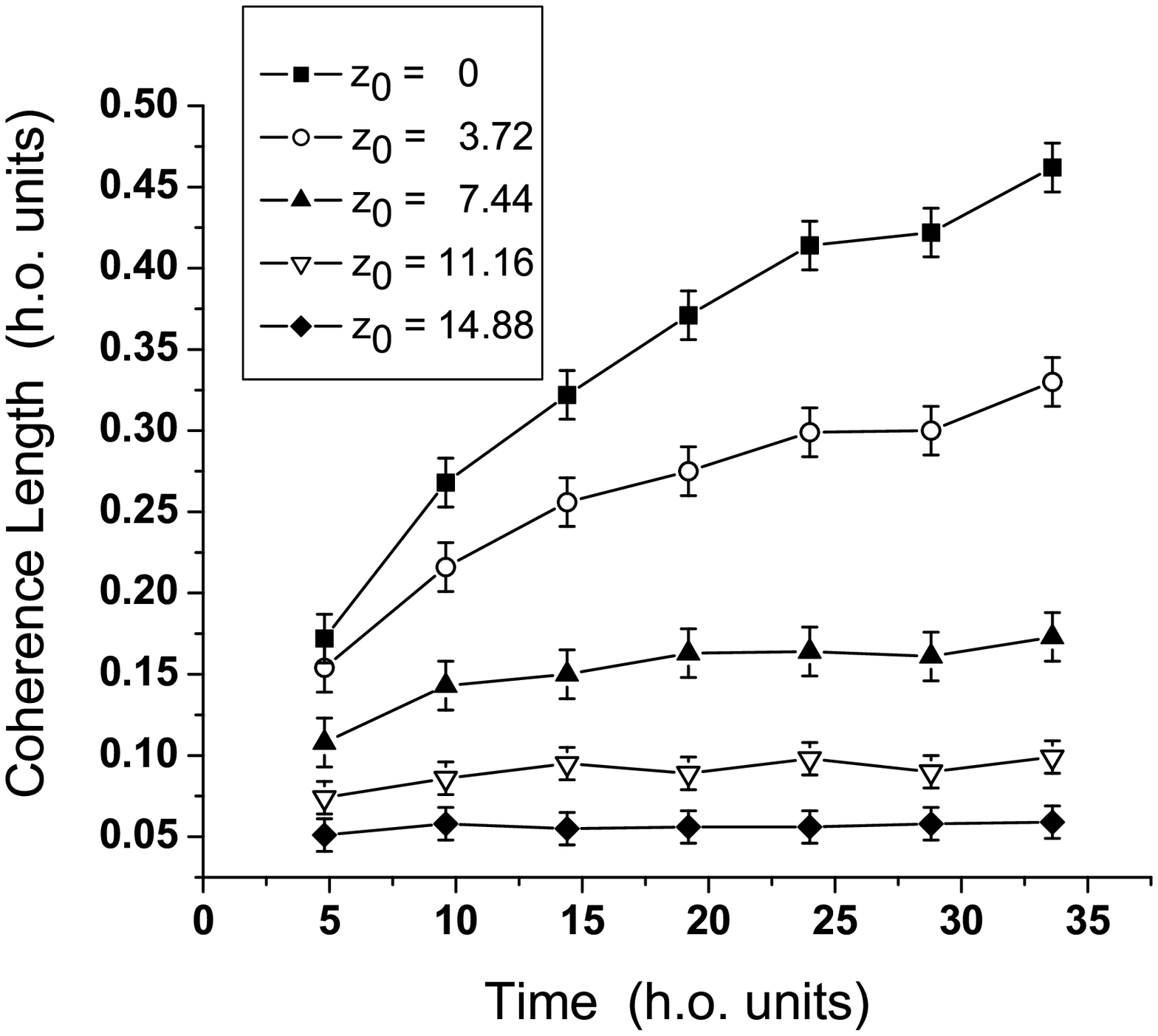,height=7.0cm} \psfig{figure=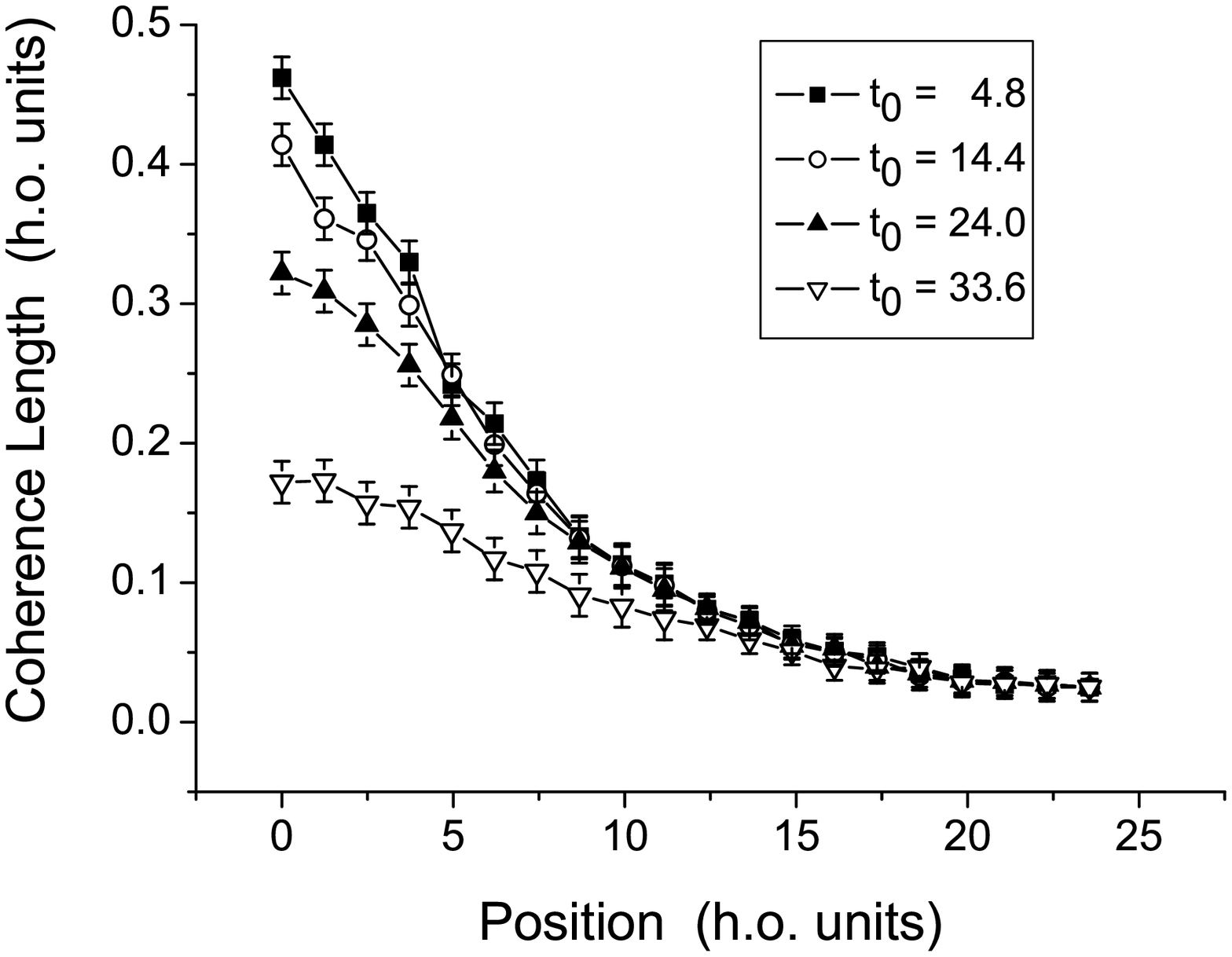,height=7.0cm}}
\caption{Behaviour of the coherence length of the trapped component without
output coupling (a) as a function of relaxation time $t_{0}$, 
at fixed points $z_{0}$ in the
trap; from top to bottom: $z_{0}/l_{z}=$ 0 (i.e. trap centre, squares), 3.72 (hollow circles),
7.44 (i.e. edge of Thomas-Fermi profile, upper triangles), 11.16 (hollow lower triangles) 
and 14.88 (rhombus). (b) $l_{coh}$ as a function of position  $z_{0}$ for different relaxation
times $t_{0}/\tau=$ 4.8 (squares), 14.4 (hollow circles), 24 (upper triangles) and 33.6
(hollow lower triangles). As a guide to the eye, neighbouring data points have been
connected by straight lines.   }
\end{figure}

\subsection{With Outcoupling}


 The coherence length of the trapped component can
decrease dramatically once the outcoupling is turned on, as a result of
the removal of coherent atoms from 
the trap centre where the Raman transition resonance condition is satisfied.
The values of the coherence length of the trapped component once steady state is
reached between pumping and outcoupling are (potentially much) lower 
than the original values
prior to outcoupling, because the atoms pumped from the surrounding thermal cloud
to compensate for the atoms lost due to the outcoupling mechanism,
do not necessarily share the same phase as the trapped atoms before their outcoupling.
 This is
shown clearly in Fig. 5(a), which also demonstrates that the
 decrease in coherence is more dramatic
at the trap centre (since atoms are mostly removed from that region) and essentially
negligible outside the Thomas-Fermi radius of the system. 
Fig. 5(a) focuses on
the regime of strong outcoupling where these effects are pronounced. In a typical
experiment, one might seek to minimize the perturbations on the trapped component
due to the outcoupling, and such a case
(corresponding to the `weak outcoupling' of Fig. 2(b)) will be further considered in Sec. 4.

\begin{figure}
\centerline{ \psfig{figure=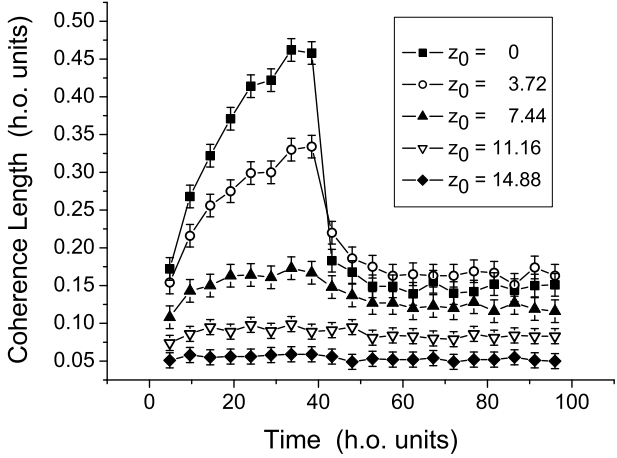,height=7.0cm} \psfig{figure=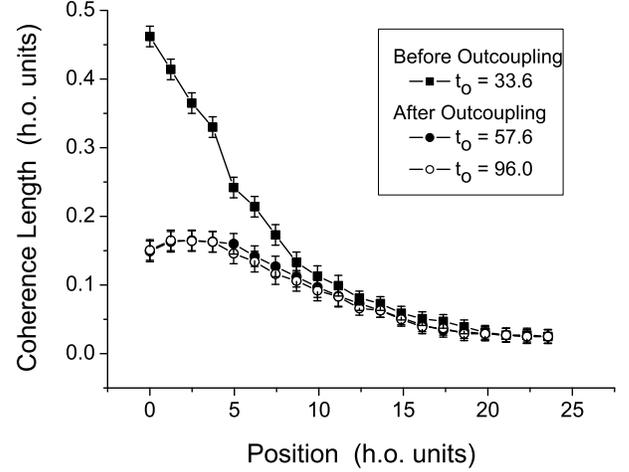,height=7.0cm}}
\caption{Behaviour of the coherence length of the trapped component with
`strong outcoupling' ($\Omega_{0}= 5 \hbar \omega_{z}$) in the limit of `large
momentum kick' $|k| = 5 l_{z}^{-1}$: (a) $l_{coh}$ versus time at points 
$z_{0}/l_{z}=$ 0 (squares),
3.72 (hollow circles), 7.44 (upper triangles), 11.16 (hollow lower triangles) and
14.88 (rhombus).
(b) $l_{coh}$ versus position $z_{0}$ at steady state (filled and hollow circles
corresponding respectively to $t_{0}/\tau=57.6$ and $96.0$). 
Outcoupling starts at $t_{0}/\tau=38.4$.
Shown also with squares
is the coherence length attained before outcoupling ($t_{0}/\tau=33.6$).}
\end{figure}

Fig. 5(b) shows the dependence of
the coherence length of the trapped component on position, once steady state has been
reached.
Apart from the rapid decrease of the coherence length due to the
outcoupling procedure, we find that
the coherence length is no longer maximum at the trap centre, but
this maximum in now slightly offset along the z-axis. 
This reflects the fact that the maximum of the quasi-condensate density is
slightly offset from its original value (as seen in Fig. 2; note that the positive z-axis
of Figs. 3-9 corresponds to the negative z-axis of Fig. 2), due to the
 asymmetry created from the mean-field interaction
between trapped and outcoupled atoms as a result of the directional
motion of the outcoupled component.

\begin{figure}
\centerline{\psfig{figure=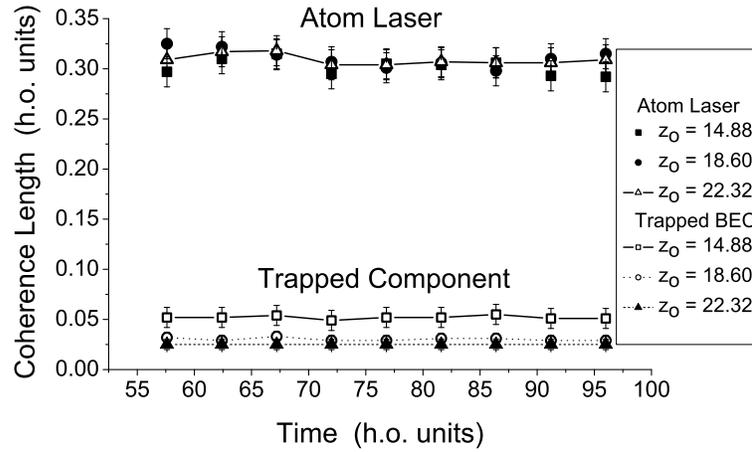,height=6.5cm}}
\caption{Coherence length of an atom laser beam as a function of relaxation time  $t_{0}$ for
three different points, located well outside the Thomas-Fermi region:
$z_{0}/l_{z}=$ 14.88 (squares), 18.6 (circles) and 22.32 (hollow triangles), with the
last data points connected by a solid line.
The corresponding values of $l_{coh}$ for the trapped
component are also plotted for comparison (using similar symbols). 
As above, $\Omega_{0}= 5 \hbar \omega_{z}$ and $|k| = 5 l_{z}^{-1}$.}
\end{figure}

\newpage
\section{Coherence Length of Atom Laser Beam}


In order to discuss the  coherence of an atom laser, we
 now perform the same type of analysis for the outcoupled component.
 Fig. 6 shows the coherence length of the atom laser beam as a function of time
for particular points along the beam, chosen such that they lie 
well outside the Thomas-Fermi region where
the interaction with the trapped quasi-condensate significantly affects its behaviour.
From this figure, we deduce that the atom laser is essentially operating under 
steady-state conditions. Fig. 7 shows the position dependence
 of the coherence length of the trapped
and outcoupled components at steady state, over the entire range of the trap. 
The coherence length
of the outcoupled component is found to change
 considerably within  the region where the trapped 
component exhibits significant coherence (ie essentially within the Thomas-Fermi radius
of the quasi-condensate), and reaches a steady-state value 
far away from the center of
the trap. On the contrary, the coherence length of the trapped component (apart from the
minor increase at small values of $|z|$ due to the asymmetry in the quasi-condensate
density profile discussed above) decreases significantly as we move away from the centre
of the trap. This shows that the outcoupled component acts as an atom laser, 
in the sense that it has a reasonably constant coherence length both in space and in time.

\begin{figure}
\centerline{\psfig{figure=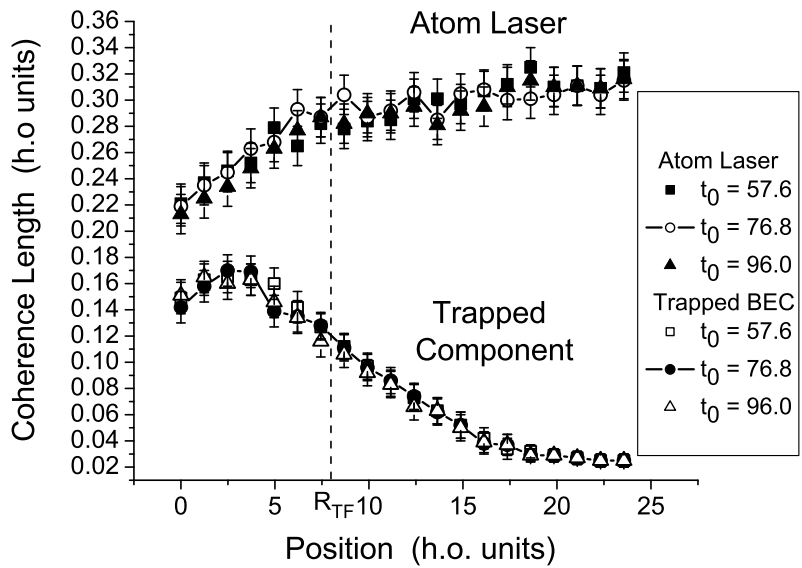,height=8.5cm}}
\caption{Steady state dependence of the coherence length of trapped (lower branch)
and outcoupled (upped branch) atoms on position 
($\Omega_{0}= 5 \hbar \omega_{z}$, $|k| = 5 l_{z}^{-1}$).
Each curve is composed of three essentially overlapping sets of values corresponding to
$t_{0}/\tau =57.6$, 76.8 and 96.0, with the intermediate data points joined by
straight lines. The dotted line indicates the position of the
Thomas-Fermi radius.}
\end{figure}

All figures shown so far focused on the case of strong outcoupling which heavily
depletes the quasi-condensate (since, in this limit, effects become more noticeable).
On the contrary, the situation of most experimental interest is likely to be 
that in which the
trapped component is only slightly perturbed by the outcoupling. This regime is
discussed in Fig. 8 which plots the position dependence of the coherence length 
of both quasi-condensate and atom laser, 
 for two different coupling strengths $\Omega_{0}$, focusing on the region of interest
sufficiently far from the trap centre.
The coherence length of the trapped component decreases rapidly towards the edges of the trap
(since there is no off-diagonal long range order in one dimension) and is larger in the
case of  weaker outcoupling, since this leads to less depletion of the
quasi-condensate (with the effect more pronounced within the Thomas-Fermi region which is
not shown here). We believe that it is for the same reasons, that we find
 a longer steady state atom laser coherence length in the case
of stronger outcoupling (see also \cite{Wiseman_Vaccaro}).

\begin{figure}
\centerline{\psfig{figure=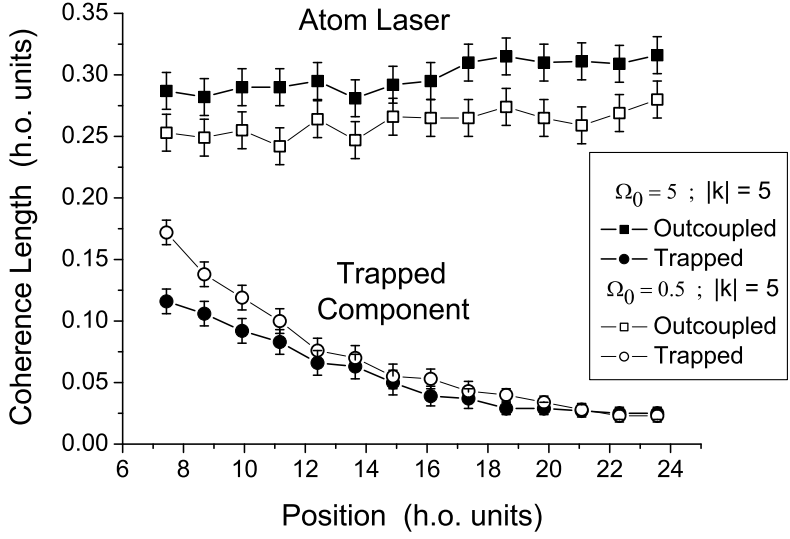,height=8.0cm}}
\caption{Steady state ($t_{0}/\tau=96.0$) coherence length of quasi-condensate 
(lower branch, denoted by circles)  and atom laser
(upper branch, denoted by squares) 
as a function of position $z_{0}$ for two different outcoupling
strengths $\Omega_{0}$, for points outside the Thomas-Fermi radius ($z_{0} > R_{TF}$):
(i) `Weak Outcoupling'  $\Omega_{0}= 0.5 \hbar \omega_{z}$ (hollow symbols)
and (ii) `Strong Outcoupling'  $\Omega_{0}= 5 \hbar \omega_{z}$ (filled symbols).
In both cases, $|k| = 5 l_{z}^{-1}$.}
\end{figure}

Finally, Fig. 9 investigates the effect of imparting different `momentum kicks' $k$ 
to the outcoupled atoms, at a  
constant outcoupling rate $\Omega_{0}$. The coherence length of the trapped 
component in the region $z_{0} < 12 l_{z}$ is found to be 
somewhat larger in the limit of smaller
momentum kicks. A possible interpretation, may be the following: 
since the outcoupled atoms leave the
central `interaction region' at a slower rate, the non-markovian nature of the 
mean field interactions between the
two components \cite{Phen3} ensures that there is always a larger fraction of 
trapped atoms in this case,
and hence more coherence within the trapped component. On the other hand, the
coherence length of the atom laser beam is significantly increased for larger momenta,
due to the increased speed of propagation of the outcoupled atoms in the waveguide.

\begin{figure}
\centerline{\psfig{figure=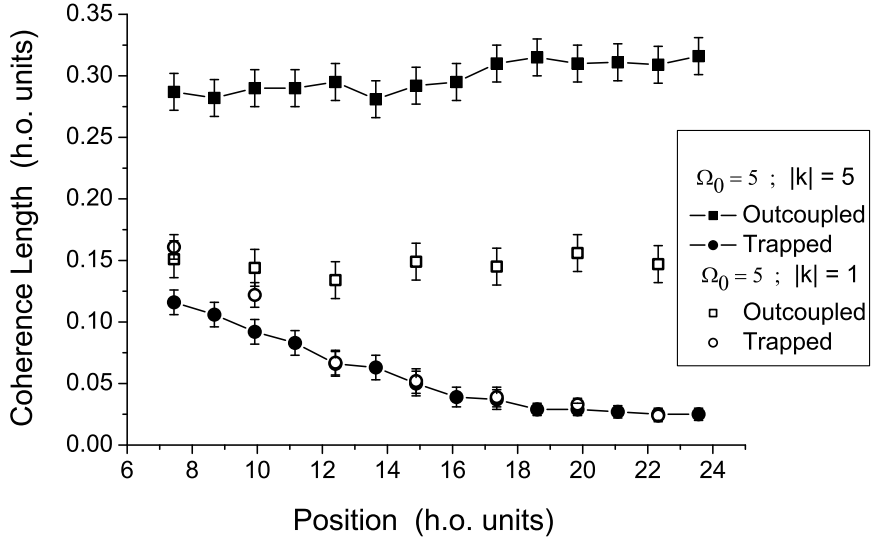,height=8.0cm}}
\caption{Steady state ($t_{0}/\tau=96.0$) coherence length of  quasi-condensate 
(denoted by circles)  and atom laser
(denoted by squares)      
 outside the Thomas-Fermi radius as a function of position,
for two different values of imparted momentum kicks to the outcoupled atoms,
in the limit of `strong outcoupling' $\Omega_{0}= 5 \hbar \omega_{z}$:
Filled symbols are the same as in Fig. 8 and indicate 
`large momentum kick' $|k| = 5 l_{z}^{-1}$ ($v \sim 1.2 mm/s$), whereas 
hollow symbols  stand for `small momentum kick' $k = l_{z}^{-1}$ ($v \sim 0.24 mm/s$)}
\end{figure}

\section{Conclusions}


The coherence properties of an atom laser beam are still only partially understood,
 despite
significant work done on this field both experimentally and theoretically 
over the past few years. In this paper, we have focused
on a one-dimensional atom laser beam in a tightly-confining waveguide, and we have shown
that it can indeed maintain a reasonably constant coherence length in both space
and time, in the region sufficiently far away from
the centre of the trap (where the perturbations due to interactions with the
trapped coherent component are minimized). 
Our system was assumed to be kinematically one-dimensional, and thus 
our analysis completely suppressed the effect of the
transverse directions of the waveguide in which the atom laser propagates.
In three-dimensional systems, the effects of the  transverse modes
can become significant, as discussed in \cite{Aspect_AL,Busch_AL}.
Our values for the coherence length of the atom laser beam are considerably
smaller than the corresponding three-dimensional systems, because it is well-known 
that phase fluctuations in low-dimensional systems can be strongly enhanced, 
leading to the
appearance of quasi-condensation (that is, condensation with rapidly varying phase
over the size of the system), as opposed to true condensation which arises in 
three-dimensional systems. In principle, this `defect' can be cured, 
by working in that region of the phase diagram,
 where one essentially obtains `true BEC' \cite{Phase_Diagram}. 
A further limitation of the
approach discussed in this paper is that for computational reasons, 
the thermal cloud was treated clasically, 
which means that its size was
slightly overestimated, leading to a minor
underestimate in the coherence properties of our system. Nonetheless, simulations of the
full quantum theory have shown this effect to be a rather minor correction,
at least for the parameters considered here \cite{Low_D_Paper}.

Notwithstanding the above remarks, we consider our current contribution as 
yet another important step  in the description of the coherence properties of atom 
lasers since, to the best of our knowledge, our work is the first
qualitative and quantitative comparison of the coherence length of (quasi-)condensates
and atom lasers beyond the usual mean field treatments.
The analysis presented in this paper is based on somewhat preliminary results, and we 
intend to further investigate and support our conclusions in a subsequent publication.
Furthermore, we are currently investigating the related issues of coherence time and 
linewidth of such atom lasers for experimentally relevant parameters 
(and in the limit of almost pure condensation), and such results will be
presented elsewhere.

\vspace{1.0cm}
\begin{center}
{ACKNOWLEDGEMENTS} \\
\end{center}

I am indebted to Professor Henk Stoof for suggesting the application of the particular
stochastic field method to study the coherence properties of atom lasers, and for his
invaluable comments during the early stages of this work. I would also like to
acknowledge computational assistance from Mark Leadbeater, and the hospitality of the
Benasque Center for Science, where I had the opportunity to discuss my results with
various participants of the `Physics of Ultracold Dilute Atomic Gases' Workshop.
Funding for this work was provided by the UK EPSRC.


\end{document}